\documentclass[twoside,twocolumn, 9pt]{revtex4}
\usepackage{graphicx}% Include figure files
\usepackage{dcolumn}% Align table columns on decimal point
\usepackage{bm}% bold math
\usepackage{multirow}
\usepackage{color}
\usepackage{hyperref}
\usepackage{comment}
\usepackage{graphicx}
\usepackage{amssymb}
\usepackage{amsmath}
\usepackage{amsfonts}

\begin{document}

\title{A reentrancy of motility-induced phase separation in overdamped active Brownian particles}

\author{Hiroya Yamamoto}

\affiliation{Department of Physics, Graduate School of Science,
Tohoku University, Sendai 980-8578, Japan.}

\begin{abstract}
In a system of Self-Propelled Particles (SPPs), the combination of self-propulsion and excluded volume effects can result in a phase separation called Motility-Induced Phase Separation (MIPS).
Previous studies reported that MIPS is one of the phenomena so-called "reentrant phase separation" ,i.e., MIPS is suppressed when the Péclet number $Pe$ (dimensionless self-propelled speed) is sufficiently large.
We used a fundamental model of SPPs, i.e., overdamped Active Brownian Partcles (ABPs), to investigate the mechanism of the reentrancy of MIPS.
We expect that elucidating the conditions under which MIPS occur is important, since MIPS is a phenomenon that can occur in a wide range of SPPs systems, and the potential applications of MIPS can also be wide range.
Detailed investigation of particle motion revealed that a entire particle cluster deforms due to multiple slip deformation (known as plastic deformation in materials science).
As $Pe$ increases, the frequency of occurrence of slip-lines increases, and the particle motion becomes fluid-like.
Therefore, the shape of the cluster becomes unstable and the number of particles in the cluster decreases.
Let $\bm{f}_{LA}$ be the local spatial average of the self-propelled force generated by the particles.
The observation of the inhomogeneity in the magnitude and direction of $\bm{f}_{LA}$ shows that 
$\bm{f}_{LA}$ is large on the cluster surface and generally orients toward the cluster inside.
We determined that $\bm{f}_{LA}$ generates stress on the cluster, and it causes the multiple slip deformation.  
\end{abstract}
\maketitle

\section{Introduction}
Active matter made of self-propelled particles (SPPs) has been studied extensively in recent years.
As self-propulsion is common over a wide range of systems, SPPs can be observed in various systems from microscopic to macroscopic such as bacteria, insects, fish, animals\cite{VICSEK201271}, and vehicles\cite{RevModPhys.73.1067}.
SPPs is a non-equilibrium system in which each particle continuously consumes energy by self-propulsion. 
A combination of self-propulsion and particle interactions (attraction, repulsion, alignment, etc.) results in characteristic phenomena that are not seen in equilibrium systems.
Examples of natural systems include bacterial quorum sensing, animal swarm\cite{VICSEK201271}, and traffic jam\cite{RevModPhys.73.1067}.

Mathematical models for SPPs include Active Brownian Partcles(ABPs)\cite{PhysRevLett.108.235702, ten_Hagen_2011, Romanczuk2012}, run-and-tumble particles\cite{Cates_2012, Solon2015, D1SM01006A}, active Ornstein–Uhlenbeck particles\cite{PhysRevLett.117.038103, C4SM00665H}, etc.
Experimental realizations of SPPs can be found as Janus particles\cite{PhysRevLett.105.268302, C1SM05960B}, Quincke rollers\cite{10120318, Dommersnes2016, PhysRevX.9.031043}, etc.

In this study, we focus on Motility-Induced Phase Separation (MIPS)\cite{PhysRevLett.108.235702, annurev:/content/journals/10.1146/annurev-conmatphys-031214-014710, PhysRevLett.110.238301, PhysRevLett.110.055701, Bialké_2013}, a type of characteristic phenomenon of SPP systems.
MIPS is a phase separation resulting from a combination of self-propulsion and excluded volume effects.
It produces a coexistence between the dilute and dense phases.
Then, MIPS is believed to result from the positive feedback of following two factors:
\begin{itemize}
  \item When the local density is increased by fluctuation, the particle speed in such a region is reduced due to collisions between particles.
  \item As the particle outflow from the cluster depends on the particle speed, it becomes smaller in the region with small particle speed, which leads to an accumulation of particles in such a region.
\end{itemize}
\begin{comment}
In addition, MIPS has been reported to be a phenomenon similar to the liquid-vapor phase transition\cite{Wysocki_2014, C3SM52813H, PhysRevLett.115.098301}. 
\end{comment}

Since the coupling between self-propulsion and excluded volume effects is commonly observed, MIPS is also expected to be a common phenomenon over many SPP systems.
However, there are few experimental systems in which the realization of MIPS has been confirmed\cite{PhysRevX.9.031043}.
This is not because MIPS occurs only under special conditions, but rather
because it is difficult to distinguish between MIPS and other phase separation phenomena, especially when there is attraction between particles.

It is reported that MIPS is reentrant phase separation with respect to vary magnitude of the self-propelled force.
There are three types of MIPS reentrancy.
The first is the case where the repulsion force of the inter-particle potential has a finite upper limit\cite{C3SM52469H}.
In this case, MIPS does not occur when the self-propelled force is greater than the upper limit of repulsion.
The second is when SPPs have an inertia effect\cite{PhysRevLett.123.228001}.
Even if the orientation of the particle changes slightly, the particle will strongly push away the forward particles due to inertia.
Therefore, as the self-propelled force becomes large, MIPS becomes unstable.
The third is when the mass of the particles is set to zero and there is no upper limit of repulsion (e.g., WCA potential).
This type of reentrancy has been expected to exist for some time\cite{doi:10.1073/pnas.1116334109} and was recently confirmed by numerical simulations\cite{Su2023}.
Kinetic analysis, in which the inflow and outflow of particles on the cluster surface are balanced at a steady state, showed that the number of particles in the clusters is reduced by evaporation when the Péclet number is very large ($Pe=\mathcal{O}(1000)$).
It was also shown that hard inter-particle potentials suppress evaporation.

The purpose of this study is to clarify the conditions under which MIPS occurs through an investigation of MIPS reentrancy.
We expect that elucidating the conditions is important, since MIPS is a phenomenon that can occur in a wide range of SPPs systems, and the potential applications of MIPS can also be wide range.

\section{Models and calculation methods}
As a model for SPPs, we used ABPs moving on a two-dimensional plane.
The equation of motion is defined by the following overdamped Langevin equation.
\begin{align}
  \nonumber
  \frac{d\bm{r}_i(t)}{dt} &=
  v_0\bm{n}_i(t) 
  -\frac{1}{\gamma}\sum_j\nabla U(|\bm{r}_i-\bm{r}_j|)\\
  \label{equ: EOM1}
  &~~~+\sqrt{2D_T}\bm{\xi}_i(t)\\
  \label{equ: EOM2}
  \frac{d\theta_i}{dt} &=
  \sqrt{2D_R}\eta_i(t)
\end{align}
Here, $\bm{r}_i=(x_i, y_i)^{\mathsf T}$ and $\theta_i$ are the position and orientation of $i$-th particle, respectively, $v_0$ is the speed of self-propulsion, $\bm{n}_i=(\cos\theta_i, \sin\theta_i)^{\mathsf T}$ is a unit vector representing the direction of self-propulsion of $i$-th particle, $\gamma$ is the drag coefficient and $-\sum_j\nabla U(|\bm{r}_i-\bm{r}_j|)$ is the sum of potential forces between particles acting on $i$-th particle.
$D_T$ and $D_R$ are the translational and rotational diffusion coefficients, respectively. 
$\bm{\xi}_i=(\xi_{ix}, \xi_{iy})^{\mathsf T}$ and $\eta_i$ are the unit variance and zero mean white Gaussian noises.

The inter-particle potential $U$ is defined as the WCA potential\cite{10.1063/1.1674820} 
\begin{align}
  U(r) = 
  \left\{
  \begin{array}{ll}
  4\epsilon 
  \left[
    \displaystyle
    \left(\frac{\sigma}{r}\right)^{12}
    -\left(\frac{\sigma}{r}\right)^{6}
  \right]+\epsilon & (r<2^{1/6}\sigma) \\
  0 & (r\ge2^{1/6}\sigma)
  \end{array}
  \right. .
\end{align}
Here, $\epsilon$ and $\sigma$ are potential parameters.
Then, we defined the particle diameter as $\sigma_d = 2^{1/6}\sigma$.

The temperature of the system is $T$, and the Einstein relation\cite{Ken2010}
\begin{align}
  \label{equ: Ein relation}
  D_T = \frac{k_BT}{\gamma}
\end{align}
is assumed to be satisfied.
The Reynolds number is assumed to be small so that the Stokes-Einstein-Debye relation\cite{Debye1929, Berne2000, 10.1063/1.470495}
\begin{align}
  \label{equ: Ein-sto relation}
  D_T = \frac{\sigma_d^2 D_R}{3}
\end{align}
is satisfied.
Eqn (\ref{equ: Ein-sto relation}) implies that the random torque of rotation, like the random force of translation, is due to collisions of solvent particles.
It should be noted that eqn (\ref{equ: Ein-sto relation}) is not always fulfilled.
This is because there are various types of rotational motion of SPPs, such as the run-and-tumble motion of bacteria or the changing direction of vehicle.
There are many studies that treat $D_T$ and $D_R$ as independent parameters\cite{10.1063/5.0040141, PhysRevX.9.031043, Solon2015}.
In this study, however, eqn (\ref{equ: Ein-sto relation}) is assumed to reduce the number of independent parameters.

We defined the units of length, time and temperature as $\sigma_d$, $1/D_R$ and $\epsilon/k_B$, respectively. 
Using these units and two equations eqn (\ref{equ: Ein relation}) and eqn (\ref{equ: Ein-sto relation}), the two independent nondimensional parameters in the equations of motion are a Péclet number $Pe = v_0\sigma_d/(3D_T)$ and a dimensionless temperature $T'=3D_Tk_BT/\epsilon$.
The dimensionless equations of motion and the WCA potential are shown in the Electronic supplementary information (ESI, It is at the bottom of this preprint).
The magnitude of dimensionless self-propelled force (i.e., dimensionless self-propelled speed) is $Pe/3$.
Since $T'$ only determines the hardness of the WCA potential, it is fixed at $T'=0.05$.
In the following, all quantities are assumed to be dimensionless using the above units.
For simplicity, the symbols of the variables were left unchanged as those before the nondimensionalization ($\bm{r}_i$, $t$, etc.).

Let the system be a square with one side length $L$ and the number of particles be $N$.
The number density of overall system is $\rho_{tot}=N/L^2$.
In addition to $Pe$, the area fraction of the ABPs $\Phi=\pi\rho_{tot}/4$ and the number of particles $N$ (indicating finite-size effects) are the other independent parameters of the system.

The simulations were performed with the standard molecular dynamics method on eqn (\ref{equ: EOM1}) and eqn (\ref{equ: EOM2}).
In the initial state, the positions of all particles are random and homogeneous, and the orientations are random and uniform.
The time integration was performed up to $1000.0$ (unit time), when the system is confirmed to reach its steady state.
Periodic boundary conditions were used in both $x$- and $y$-directions.
Three parameters $Pe$, $\Phi$, and $N$ were varied in the intervals $50\le Pe \le1000$, $0.5\le\Phi\le0.8$, and $1024\le N\le8192$, respectively.
Time step was $\Delta t=10^{-5}$ for $0 \le Pe < 100$ and $\Delta t=10^{-6}$ for $100 \le Pe < 1000$.
In the actual calculation, we used HOOMD-blue\cite{ANDERSON2020109363} (i.e., a general-purpose particle simulation toolkit).

The number of samples simulated by different initial conditions, which are used for the following analyses, is shown in the ESI.

\section{How to determine the maximum cluster}
\begin{figure}[b]
  \begin{center}
    \includegraphics[width=8cm]{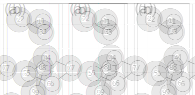}
    \caption{An example of the graph corresponding to the positions of the particles.
    (a)The particle positions. 
    Blue circles indicate the particles and numbers inside the circles indicate the indices of the particles.
    (b)The graph corresponding to the positions of the particles shown in (a).
    White circles indicate vertices and black line segments connecting vertices indicate edges.
    \label{fig: ss2graph}}
  \end{center}
\end{figure}

The agglomerated particles produced by MIPS are called clusters, but its precise definition was rarely given in the past studies.
%The agglomerated particles produced by MIPS are called clusters, but its precise definition was not given in the past studies.
In this section, we will define the maximum cluster for the analysis in subsequent sections.
The maximum cluster is identified using network theory\cite{estrada2013, grinberg2023} presented below.
Let us assume the index $i$ of a particle as the vertex of a graph, and the set of vertices as $V=\{1,2,...,N\}$.
The radial distribution function(RDF) of particles is computed by the positions $\{\bm{r}_i\}$.
We defined $r_{1 \rm st}$ as the first peak position of RDF, and also defined the threshold distance $r^*$ as $r^*=(r_{1\rm st}+\sqrt{3}r_{1\rm st})/2$.
Inside clusters, the particles are arranged in a triangular lattice, so the $r^*$ is the midway between the first and second peak of RDF.
For any two distinct vertices $i, j\in V, ~(i\neq j)$, an edge $\{i, j\}$ of the graph is defined if $|\bm{r}_i - \bm{r}_j| \le r^*$ is satisfied.
Let $E$ be the set of edges.
The graph $G$ is defined as the ordered pair $G=(V, E)$.
A specific example of determining $V$ and $E$ for the positions of particles is shown in Fig. \ref{fig: ss2graph}.
In this example, $V$ and $E$ are 
\begin{align}
  V &= \{1,2,3,4,5,6,7,8,9,10\},\\
  \nonumber
  E &= \{
    \{2,3\},
    \{4,5\},\{4,6\},\{4,7\},\{5,6\},\{6,7\},\\
    &~~~~~~~~\{8,9\},\{9,10\}
    \} ,
\end{align}
respectively, and the graph is visualized as Fig. \ref{fig: ss2graph}(b).

In network theory, a path is defined as a sequence of vertices, where there are two edges between each vertex and the vertices before and after it.
$V$ is partitioned into several subsets depending on whether a path exists or not between the vertices.
We assumed the maximum cluster as the largest subset of the partitioned $V$ and denoted it by the symbol $V_{cl}$.
In this study, because $N$ is small ($1024\le N\le8192$), there is only one large cluster in the system, and it is the maximum cluster.
In the example in Fig. \ref{fig: ss2graph}(b), $V$ is partitioned into
\begin{align}
  V &= \{1\}\cup\{2,3\}\cup\{4,5,6,7\}\cup\{8,9,10\}
\end{align}
and $V_{cl}$ is
\begin{align}
  V_{cl}=\{4,5,6,7\} .
\end{align}

In the same way as partitioning $V$, $E$ is also partitioned.
If the edges do not exist, the paths do not exist either.
Let $E_{cl}$ be one of the subsets of the partitioned $E$ that is formed by the vertices in $V_{cl}$.
In the example in Fig. \ref{fig: ss2graph}(b), $E$ is partitioned into
\begin{align}
\nonumber
  E &= 
  \{\{2,3\}\}\cup
  \{\{4,5\},\{4,7\},\{5,6\},\{6,7\}\}\\
  &~~~~~\cup\{\{8,9\},\{9,10\}\}
\end{align}
and $E_{cl}$ is
\begin{align}
  E_{cl} &= 
  \{\{4,5\},\{4,7\},\{5,6\},\{6,7\}\}.
\end{align}

Also, $G_{cl} = (V_{cl}, E_{cl})$ is called the maximal connected (paths exist between any two vertices) subgraph of the graph $G$.

\section{Coexistence densities}
Let us consider a parameter setting $(Pe, \Phi, N)$ where MIPS occurs in steady state.
In such a steady state, we performed the Voronoi tessellation\cite{Voronoi1908} on the particle positions $\{\bm{r}_i\}$.
The Voronoi tessellation partitions the system into $N$ Voronoi cells.
Since each cell contains only one particle, the cell is identified by $i$, the same as the particle index.
We defined $s_i$ as the area of the Voronoi cell $i$ and $\varphi_i=1/s_i$ as the number density (the Voronoi-based density) of the cell $i$.
Let $f(\varphi)$ be the distribution function of the Voronoi-based density $\varphi$.
When MIPS occurs, $f(\varphi)$ has a bimodal distribution.
The two peaks of $f(\varphi)$ are caused by the dilute and dense phases.

Here, there is a point to note about the Voronoi-based density.
The number density used in ordinary statistical physics is the grid-based density.
It is obtained by partitioning the system into grids and dividing the number of particles in each grid by the area of the grid.
Let $g(\rho)$ be the distribution function of the grid-based density $\rho$.
In general, $g$ and $f$ are different.
The expected value of $\rho$ is $N/L^2$, but the expected value of $\varphi$ is not $N/L^2$\cite{ESI}.
However, it is possible to obtain $N/L^2$ from $f(\varphi)$.
Since the expected value of $1/\varphi$ is equal to the expected value of the Voronoi cell area ($=L^2/N$), the following relationship holds.
\begin{align}
  \label{equ: varphi to rho mean}
  \int_0^{\infty} d\rho~\rho g(\rho) = \left[ \int_0^{\infty} d\varphi~\frac{f(\varphi)}{\varphi} \right]^{-1}
\end{align}

In light of the above relationship, we will explain how to obtain the coexistence densities for MIPS.
From the simulation data, $f(\varphi)$ is obtained. 
It has two peaks corresponding to the dilute and dense phases.
Therefore, we assumed that $f(\varphi)$ is described by a superposition of two functions, each of which can be fitted by two functions $f_g$ and $f_\ell$.
The specific forms of $f_g$ and $f_\ell$ are
\begin{align}
  f_g(\varphi; C, \mu, \lambda)
  &= C\exp
  \left(
    -\frac{(\ln\varphi-\mu)^2}{2\lambda^2}
  \right),\\
  f_\ell(\varphi; D, \varphi_0, a)
  &=
  \frac{D}{|\varphi-\varphi_0|^a}
  \exp
  \left(
    -\frac{1}{|\varphi-\varphi_0|}
  \right) ~~~(\varphi<\varphi_0),
\end{align}
where $C$, $\mu$, $\lambda$, $D$, $\varphi_0$ and $a$ are the fitting parameters.
$f_g$ is a function called the Log-Normal Distribution (LND).
Because LND increases rapidly and decays slowly, we can fit LND to the peak of $f(\varphi)$ caused by the dilute phase with good precision.
LND has been commonly used in studies of various dilute systems\cite{Yang2002,Over2006,Usui2018,Dornan2024}.
On the other hand, $f_\ell$ is a new function introduced in this study.
Since the main peak of $f(\varphi)$ that corresponds to the dense phase increases slowly and decays rapidly, we used the fitting function with the same property.
Specific fitting results are shown in the ESI.
By eqn (\ref{equ: varphi to rho mean}), the expected values of $\rho$ in the dilute phase $\rho_g$ and the dense phase $\rho_\ell$ are
\begin{align}
  \label{equ: rho1 rho2}
  \rho_g &= \left[ \int_0^{\infty} d\varphi~\frac{f_g(\varphi)}{\varphi} \right]^{-1},\\
  \rho_\ell &= \left[ \int_0^{\infty} d\varphi~\frac{f_\ell(\varphi)}{\varphi} \right]^{-1}.
\end{align}
$\rho_g$ and $\rho_\ell$ are the coexistence densities.

There are an advantage and a disadvantage of using the Voronoi-based density instead of using the grid-based density.
The advantage is that there is no need to adjust the grid size.
This is because the Voronoi tessellation partitions the system automatically.
The disadvantage is that the fitting function of $f(\varphi)$ is unknown.
In fitting for $g(\rho)$, two Gaussian functions are often used\cite{PhysRevE.98.030601}.
On the other hand, there is no fitting function that is conventionally used for $f(\varphi)$ (in particular $f_\ell$).

\begin{figure}[b]
  \begin{center}
    \includegraphics[width=8cm]{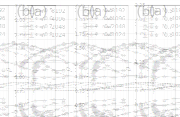}
    \caption{Coexistence densities.
    The downward and upward triangular symbols correspond to 
    $\rho_g$ and $\rho_\ell$, respectively.
    The dotted lines show the average values of $\rho_g$ and $\rho_\ell$.
    (a)The result for varying $Pe$ and $N$ with $\Phi=0.6$ fixed.
    (b)The result for varying $Pe$ and $\Phi$ with $N=8192$ fixed.
    The correspondence between colors and parameters is as shown in the legend.
    \label{fig: rho_g and rho_ell}}
  \end{center}
\end{figure}

The results for $\rho_g$ and $\rho_\ell$ calculated with different parameters are shown in Fig. \ref{fig: rho_g and rho_ell}.
As $Pe$ increases from $Pe=0$, $\rho_\ell-\rho_g$ once increases and turns to decrease when $Pe$ is sufficiently high.
This shows that MIPS is suppressed when $Pe$ is sufficiently high and means that MIPS is reentrant.
In addition, the reentrancy weakens as $N$ increases.

\begin{figure}[b]
  \begin{center}
    \includegraphics[width=8cm]{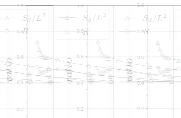}
    \caption{Lever rule.
    The two ratios $S_{cl}/L^2$ and $R$ in the case where $Pe$ is varied with $(\Phi, N)=(0.6, 8192)$ fixed.
    \label{fig: ratio}}
  \end{center}
\end{figure}

There are two additional comments on the coexistence densities.
The first is that these coexistence densities have a finite-size effect.
Let us consider the case $(Pe, \Phi, N)=(1000, 0.85, 1024)$.
$\rho_{tot}$ is approximately $1.08$, which is higher than the value of $\rho_\ell$ shown in Fig. \ref{fig: rho_g and rho_ell}(a), but MIPS occurs in the steady state.
This means that $\rho_\ell$ for $\Phi=0.6$ and $\rho_\ell$ for $\Phi=0.8$ do not match.
In Fig. \ref{fig: rho_g and rho_ell}(b) do not show this tendency.
Therefore, we judged it is the finite-size effect, which disappears as $N\to\infty$.
Second, these coexistence densities are expected to satisfy the lever rule\cite{PhysRevX.9.031043, PhysRevE.97.020602, PhysRevLett.125.168001} as well as the equilibrium system.
Let $S_{cl}$ be the total area of the Voronoi cells that the cell indices belong to $V_{cl}$.
\begin{align}
  S_{cl} = \sum_{i\in V_{cl}}s_i
\end{align}
Then, let the ratio $R$ be
\begin{align}
  R = \frac{\rho_{tot}-\rho_g}{\rho_\ell-\rho_g}.
\end{align}
In the case where $Pe$ is varied with $(\Phi, N)=(0.6, 8192)$ fixed, $S_{cl}/L^2$ and $R$ are obtained as in Fig. \ref{fig: ratio}.
The maximum cluster defined in this study is regarded to include both the dense phase and the interface region.
Therefore, $S_{cl}/L^2$ tends to take a higher value than $R$ in Fig. \ref{fig: ratio}.
From Fig. \ref{fig: ratio}, we expected that the coexistence densities satisfy the lever rule.

\section{Determination of the inside of the maximum cluster and calculation mean square displacement}
\begin{figure}[b]
  \begin{center}
    \includegraphics[width=8cm]{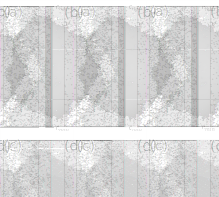}
    \caption{Some examples of CFCC calculation results.
    (a) and (b) are visualizations of the values of CFCC.
    Colored circles indicate particles contained in the maximum cluster and gray circles indicate other particles.
    The relationship between the values of centrality and colors is as shown in the colormap.
    (c) and (d) are visualizations of $V_b$.
    Orange circles indicate particles contained in $V_b$ and gray circles indicate other particles.
    \label{fig: cfcc}}
  \end{center}
\end{figure}

We focused on the particles contained inside the maximum cluster and calculated the Mean Square Displacement(MSD) of the particles in the clusters.
First, we will explain how to determine the inside of the maximum cluster.
The following determination method is just technical, so we will focus on explaining the results.
We used a concept of network theory\cite{estrada2013} to determine the inside of the maximum cluster.
It is a mapping from the vertex to a real number zero or more $c:V\to \mathbb{R}_0^{+}$, and it has various definitions for different purposes.
In this study, we used the centrality called Current-Flow Closeness Centrality(CFCC)\cite{10.1007/978-3-540-31856-9_44}.
We defined the weight of the edge $\{i,j\}~\forall i,j\in V_{cl}$ as the inverse of the distance between the particles $1/|\bm{r}_i-\bm{r}_j|$ for the calculation of CFCC.
Some examples of calculation results are shown in Figs. \ref{fig: cfcc}(a) and (b).
we can see the values of CFCC are higher toward the center of the maximum cluster and they correspond to the geometry of the cluster.

We defined the inside of the maximum cluster as the set of vertices which the values of CFCC are greater than the threshold $c^*$ (defined below).
For a bimodal $f(\varphi)$, let $\varphi_1$ and $\varphi_2$ be the positions of the peaks ($\varphi_1<\varphi_2$).
Then, we defined $c^*$ as
\begin{align}
  V' &= \{i|\forall i \in V_{cl}, \varphi_i > \varphi_2\}\\
  c^* &= \frac{1}{|V'|}\sum_{i\in V'}c(i).
\end{align}
Here, $|V'|$ is the number of elements in $V'$.
$c^*$ is the mean of the CFCC values of the vertices that the values of the Voronoi-based density are greater than $\varphi_2$.
Using $c^*$, the inside of the maximum cluster $V_b$ is defined as
\begin{align}
  V_b &= \{i|\forall i \in V_{cl}, c(i) > c^*\}.
\end{align}
Examples of the calculation results of $V_b$ are shown in Figs. \ref{fig: cfcc}(c) and (d).
In this study, $c^*$ is defined so that the width of the region between the surface and the inside of the maximum cluster is several times the particle diameter (about $5.0$ or more).

\begin{figure}[b]
  \begin{center}
    \includegraphics[width=8cm]{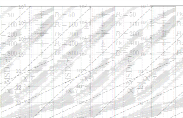}
    \caption{MSD of the particles contained in $V_b$.
    $Pe$ is varied with $(\Phi, N)=(0.6, 8192)$ fixed.
    The solid lines with the symbols are MSD.
    The vertical dashed lines indicate $\tau=5\times3/Pe$.
    The relationship between the color and $Pe$ is shown in the legend.
    The two black dashed lines are curves proportional to $\tau$ and $\tau^2$, respectively.
    \label{fig: msd}}
  \end{center}
\end{figure}

Let us assume that $Pe$ varies with $(\Phi, N)=(0.6, 8192)$ fixed.
Let $t_0$ and $t_1,~(t_0 < t_1)$ be two times in steady state, and let $\tau$ be $\tau=t_1-t_0$.
Some MSD of the particles contained in $V_b$ is calculated as shown in Fig. \ref{fig: msd}.
They show plateau-like behavior.
This is consistent with previous studies reporting that MIPS cluster is fragile glass\cite{10.1063/5.0040141}.
Since the plateau-like behaviors end about $\tau\sim3/Pe$\cite{ESI}, the MSD is proportional to $\tau^2$ at $\tau=5\times(3/Pe)$ as shown in Fig. \ref{fig: msd}.
This is not due to the effect of the particles leaving the cluster for the following two reasons.
(1) An ABP without collision and rotation propels $1.0$ (unit length) on average per $3/Pe$ (unit time).
(2) The width of the region between the surface and inside of the maximum cluster is about $5.0$ or more (by definition of $c^*$).
In the previous study\cite{PhysRevLett.110.055701}, where $Pe$ varies in $0 \le Pe \le 150$, the plateau slopes decrease as $Pe$ increase.
However, we could not observe such tendency in the present study.
In Fig. \ref{fig: msd}, the plateau slope for $Pe=800$ is greater than that for $Pe=400$, or the plateau disappears for $Pe=800$.
This behavior of the MSD means that the motion of the particles inside the cluster is fluid-like if $Pe$ is sufficiently large, and the detail of the particle motion will continue to be investigated in the next section.

\section{Particle motion in the cluster}
\begin{figure}[b]
  \begin{center}
    \includegraphics[width=8cm]{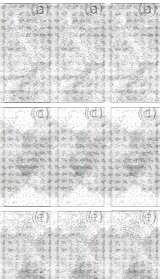}
    \caption{The displacements of the particles during $\tau^*$.
    The displacement orientations for each of the particles are shown in (a), (c), and (e).
    the colored dots are the particles.
    (a) for $Pe=100$, (c) for $Pe=400$, and (e) for $Pe=800$.
    The mean displacement orientations for each of the grids are shown in (b), (d), and (f).
    (b) for $Pe=100$, (d) for $Pe=400$, and (f) for $Pe=800$.
    The arrows are the mean displacement vectors for each grid (total number of grids is $14\times14$).
    Since $Pe>100$ and $\tau^*$ is small, the arrows are the streamlines.
    The colors of the particles and arrows correspond to the orientation $\theta$ values as shown in the colormap below the figure.
    The transmittance of the arrow is proportional to the number density of particles in the grid.
    $(\Phi, N)$ is set to $(\Phi, N)=(0.6, 8192)$.
    \label{fig: md}}
  \end{center}
\end{figure}

For studying the motion of particles in the cluster, we focused on the specific timescale $\tau^*=5\times3/Pe$.
On this timescale, the motion of the particles is ballistic.
The displacements of the particles during $\tau^*$ are shown in Fig. \ref{fig: md}.
In Fig. \ref{fig: md}, the spatial velocity correlations and vortices are observed as in previous studies\cite{D0SM02273J, Szamel_2021}.
In this study, we focused on the fact that the strongly correlated regions are divided into several domains, and the domain boundaries are straight.
We expected that the motions of each domain contribute to the MSD.

\begin{figure}[b]
  \begin{center}
    \includegraphics[width=8cm]{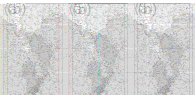}
    \caption{The positions of the defects in the cluster.
    Red and blue dots indicate particles with coordination number of 5 and 7, respectively. 
    Gray dots indicate other particles.
    (a)for Pe=100 and (b)for Pe=400 with $(\Phi, N) = (0.6, 8192)$.
    \label{fig: dislocation}}
  \end{center}
\end{figure}

To clarify the mechanism of the domain formation, we investigated the motions of the defects inside the clusters.
We created the Delaunay diagram\cite{1573105974067499264} by the particle positions $\{\bm{r}_i\}$ and assumed it to be the graph.
Inside the cluster, the particles arrange in a triangular lattice, so the number of edges connected to a vertex (i.e., the coordination number) is basically $6$, but it deviates from $6$ when there is a defect.
Basically, two particles with coordination number $5$ and $7$ appear in pairs ($5$-$7$ pairs) at a defect.
In exceptional cases, triplets with coordination number $5$, $5$, and $8$ may appear, but they are ignored in this study because their occurrence probability is quite lower than the $5$-$7$ pairs.
The positions of the pairs in the cluster are shown in Fig. \ref{fig: dislocation}.
The coordination number of particles on the cluster surface also deviate from $6$, and these particles are also ignored.

\begin{figure*}
  \begin{center}
    \includegraphics[width=18cm]{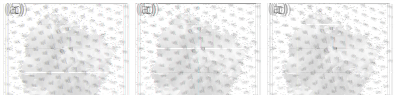}
    \caption{The relationship between the dislocation and the straight boundary.
    (a)The same figure as in Fig. \ref{fig: md}(b). The boundary we focus on is enclosed by red frame.
    (b), (c) and (d)The circles indicate particles, whose diameters are reduced for visibility.
    The red and blue circles are particles with coordination numbers 5 and 7, respectively.
    The line segments show the trajectory of each particle during $\tau$,
    and the line color corresponds to the orientation of displacement as in (a).
    \label{fig: slipline}}
  \end{center}
\end{figure*}

In a statistical study of defects in the crystalline and hexatic phases\cite{D1SM01411K}, these pairs are identified with the dislocations, which are classified into several types.  One type creates vacancies, another creates grain boundaries, and the other type is called free dislocation in the study\cite{D1SM01411K}.
In this study, we focused on free dislocations that can move within a cluster.
The relationship between the movement of the dislocation and the straight boundary between the two domains is shown in Fig. \ref{fig: slipline}.
From Fig. \ref{fig: slipline}, we can see that the trajectory of the dislocation movement coincides with the domain boundary.
This phenomenon is analogous to the fact that the trajectory of a dislocation movement in a crystal forms a slip-line\cite{Jastrzebski1977}.
We created several videos that show movements of dislocation within cluster.
We also confirmed that multiple slip deformation occurs within cluster, however, lengths of each slip are short, about the diameter of the particle.
In addition, the videos show that the dislocations are aligned on a curve and form a grain boundary.
The grain boundary is also observed in Fig. \ref{fig: dislocation}.

From Fig. \ref{fig: md}, two adjacent domains can move in exactly opposite directions to each other by slip, so a entire cluster does not translate like a rigid body.
A rigid rotation of the entire cluster has not occurred, because any axes of rigid rotation are not observed in the streamlines.
Furthermore, we examined the motion of the particles on a time scale $10\times(3/Pe)$ longer than $\tau^*$.
The entire cluster deforms significantly on this time scale\cite{ESI}.
Therefore, we judged that effects of rigid translation and rotation of the cluster on the MSD are small.
The reason that plateau hardly appears in the MSD for $Pe=800$ as shown in Fig. \ref{fig: msd} is the slip-lines increase as $Pe$ increases and the motions of the particles become fluid-like.

\section{Discussion}
The previous section revealed that the cluster is divided into multiple domains, each of which moves in a different direction.

\begin{figure}[b]
  \begin{center}
    \includegraphics[width=8cm]{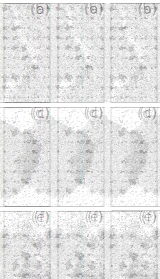}
    \caption{The particle orientations $\{\bm{n}_i\}$ and $\bm{f}_{LA}$.
    (a), (c), and (e) show the particle orientations. 
    (a)for $Pe=100$, (c)for $Pe=400$ and (e)for $Pe=800$.
    (b), (d), and (f) shows the orientation of $\bm{f}_{LA}$. 
    (b)for $Pe=100$, (d)for and (f)for $Pe=800$.
    Here, $\bm{f}_{LA}$ is average value of self-propelled forces within each grid (total number of grids is $10\times10$).
    The size of the arrow is proportional to the magnitude of the $\bm{f}_{LA}$ of the grid.
    The colormap below the figure shows the relationship between the color and the orientation.
    In all six figures, $(\Phi, N)$ is set to $(\Phi, N)=(0.6, 8192)$.
    \label{fig: orient}
    }
  \end{center}
\end{figure}

We defined $\bm{f}_{LA}(\bm{r})$ as a local spatial average of the self-propelled forces.
Figure \ref{fig: orient} shows some examples of the particle orientations $\{\bm{n}_i\}$ and $\bm{f}_{LA}$.
ABPs that have just collided cluster surface from outside tend to propel toward inside of cluster.
Once ABPs come inside of cluster, they stay in cluster for some time, and their orientations become random due to rotational diffusion.
ABPs in dilute phase propel freely, and their orientations are also random.
Then the magnitude of $\bm{f}_{LA}$ is large at the cluster surface and small otherwise, as shown in Fig. \ref{fig: orient}.
Even if the orientation vectors $\{\bm{n}_i\}$ are not perfectly antiparallel to the normal vector of the cluster surface, 
the particles can still collide with the cluster. 
After the collisions, the orientations can vary due to rotational diffusion.
Therefore, $\bm{f}_{LA}$ can also tilt from the normals of the surface as shown in Fig. \ref{fig: orient}.

\begin{figure*}
  \begin{center}
    \includegraphics[width=18cm]{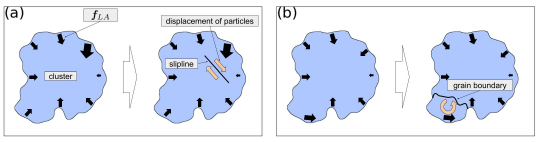}
    \caption{
    Schematic diagram shows the deformation occurs because of $\bm{f}_{LA}$ in the cluster.
    The mechanism by which $\bm{f}_{LA}$ causes the slip-line (a) and the grain boundary (b).
    In (b), tilted $\bm{f}_{LA}$ acts as a torque and creates a grain boundary.
    The domain surrounded by the boundary can rotate.
    \label{fig: cluster}}
  \end{center}
\end{figure*}

Since $\bm{f}_{LA}$ imposes a stress on the cluster, it induces deformation such as shear, rotation, extension, and compression in the cluster.
However, the types of deformation with displacements larger than the particle diameter (lattice constant) are limited.
This is because when large deformation occurs, interparticle distances of some particles significantly reduced and internal energy significantly increase.
Energy increasing because of slip deformation (it resemble shear deformation) is small, because interparticle distances decrease only around slip-lines.
Therefore, the slip occurs preferentially in the cluster.
Figure \ref{fig: cluster}(a) schematically shows that the slip-line occurs due to $\bm{f}_{LA}$.
If the cluster creates a grain boundary, particles on the boundary act as a buffer and suppress energy increasing.
Hence, the cluster can deform other than slip.
Figure \ref{fig: cluster}(b) schematically shows that tilted $\bm{f}_{LA}$ creates a grain boundary and causes a domain rotation.

We now discuss the reasons for the formation of vortices in the cluster.
Two reasons are expected.
First, a motion of particles around a slip-line or grain boundary becomes a vortex.
Let us assume that both ends of them are inside the cluster.
In this case, streamlines could be elongated circuits.
Figure \ref{fig: md}(f) shows such streamlines.
Second, a rotational motion of a domain surrounded by grain boundaries also forms a vortex.

In this study, we could not observe a specific relationship between $\bm{f}_{LA}$ and the motion of the domain.
This is because the shape of the clusters is complex and the position dependence of the stress is also complex.
Nevertheless, we expect that the effect of $\bm{f}_{LA}$ is small for larger clusters than the clusters in this study, because $\bm{f}_{LA}$ is a surface force.
This expectation is consistent with the dependence of MIPS reentrancy on $N$ as shown in Fig. \ref{fig: rho_g and rho_ell}(a).
A previous study dealing with large systems ($128000\le N, ~0\le Pe\le150$) has reported that dislocations inside a cluster move randomly\cite{PhysRevLett.110.055701}.
This means that slip-lines do not occur inside large clusters.
However, the effect of increasing $Pe$ in large clusters remains unclear.
MIPS reentrancy can create multiple clusters in the system, and the clusters can interact with each other.

\section{Conclusion}
The coexistence densities showed that the difference in number density between the dense and dilute phases $\rho_\ell-\rho_g$ becomes small when $Pe$ is sufficiently large.
This means that a sufficiently large $Pe$ suppresses MIPS.
The reason for the suppression of MIPS in this study is assumed to be that as $Pe$ increases, the motion of the particles in the cluster becomes more fluid-like and the shape of the cluster becomes unstable.
The MSD calculated for the particles inside the cluster confirms that the particles in the cluster have different motions at different time scales.
On timescales of $\tau<3/Pe$ the particles have glass-like motion and on timescales of $\tau\sim \mathcal{O}(1)\times3/Pe$ the particles have ballistic motion.
Focusing on the time scale $\tau^*\sim5\times 3/Pe$, we investigated the details of particle motion and found the slip lines and the circuit-like streamlines.
Both of these slip-lines and circuit-like streamlines were expected to be caused by local spatial averaging of self-propelled forces $\bm{f}_{LA}$.

In a previous study\cite{Su2023}, it was shown that MIPS is reentrant because an increase in $Pe$ causes the clusters to transform from active liquid to active gas.
On the other hand, in this study, we have clarified the mechanism of the transition from active solid to active liquid.
We expect that the suppression of evaporation as the potential becomes harder as shown in a previous study is caused by the increasing cost of the energy at which the slip-line is generated.

Now, we discuss possible future works.
In the present study, the Voronoi-based density was used to calculate the coexistence densities.
A shortcoming of this method is that the functional forms of the fitting functions $f_g$ and $f_\ell$ are not theoretically predicted.
Thus, theoretical predictions of the fitting functions are desired.
We assumed that the motion of the particles becomes fluid-like due to the inhomogeneity of $\bm{f}_{LA}$.
Comparison between present simulation results and those on systems with larger sizes will clarify whether our assumption is correct or not.
We also assumed that the effects of the force acting on the surface of the cluster, $\bm{f}_{LA}$, become small for large clusters.
\newpage
\paragraph*{Acknowledgments} 
The author would like to thank T.Kawakatsu for useful discussions 
and carefully proofreading the manuscript.
This work is supported by Tohoku University doctoral fellowship.
\bibliography{references_main}
\bibliographystyle{rsc}
\end{document}

\title{Supplementary information}

%\author{Hiroya Yamamoto}

%\affiliation{Department of Physics, Graduate School of Science,
%Tohoku University, Sendai 980-8578, Japan.}

\begin{abstract}
This supplementary information contains additional figures, videos and data that supplement the main text.
\end{abstract}
\maketitle

\section{Dimensionless equations of motion}
Dimensionless equations of motion are
\begin{align}
  \frac{d\bm{r}_i(t)}{dt} &=
  \frac{Pe}{3}\bm{n}_i(t) 
  -\sum_j\nabla U(|\bm{r}_i-\bm{r}_j|)
  +\sqrt{\frac{2}{3}}\bm{\xi}_i(t)\\
  \frac{d\theta_i}{dt} &=
  \sqrt{2}\eta_i(t),
\end{align}
where WCA potential $U$ is 
\begin{align}
  \displaystyle
  U(r) = 
  \left\{
  \begin{array}{ll}
  \displaystyle
  \frac{1}{3T}
  \left[
    \displaystyle
    \left(\frac{1}{r}\right)^{12}
    -2\left(\frac{1}{r}\right)^{6}
  \right]+\frac{1}{3T} & (r<1) \\
  0 & (r\ge1)
  \end{array}
  \right..
\end{align}
Here, $Pe/3$ is the dimensionless self-propelled force and $T=0.05$ is dimensionless temperature indicated in the main text.

\section{Proof of the mismatch between mean values of $g(\rho)$ and $f(\varphi)$}
Let us define $L$ as the length of one side of the system, $N$ as the number of total particles and $s_1,...,s_N$ as the areas of the Voronoi cells and $\varphi_1=1/s_1, ...,\varphi_N=1/s_N$ be the Voronoi-based densities.
Let $\bar{\rho}$ be the mean value of grid-based density $\rho$ and $\bar{\varphi}$ be the mean value of Voronoi-based density $\varphi$.
Two equations
\begin{align}
  \bar{\rho} &= 
  \left[\frac{1}{N}\sum_is_i\right]^{-1} =
  \left[\frac{L^2}{N}\right]^{-1} =
  \int_{0}^{\infty} d\rho~\rho g(\rho)\\
  \bar{\varphi} &= 
  \frac{1}{N}\sum_i\varphi_i = 
  \int_{0}^{\infty} d\varphi~\varphi f(\varphi)
\end{align}
hold.
The arithmetic and harmonic averages of the cell areas are $A$ and $H$, respectively.The following equations are satisfied.
\begin{align}
  A &= \frac{1}{N}\sum_i s_i = \frac{1}{\bar{\rho}}\\
  H^{-1} &= \frac{1}{N}\sum_i \frac{1}{s_i} = \frac{1}{N}\sum_i \varphi_i = \bar{\varphi}
\end{align}
The inequality $H \le A$ holds for $A$ and $H$, where the equality is holed only when $s_1=s_2=.... =s_N$.
Then, since
\begin{align}
  H \le A &\Rightarrow \frac{1}{\bar{\varphi}} \le \frac{1}{\bar{\rho}}\\
  &\Rightarrow \bar{\rho} \le \bar{\varphi}
\end{align}
holds, the expected values of $g(\rho)$ and $f(\varphi)$ generally do not match.

\section{Details of the fitting functions $f_g$ and $f_\ell$ for $f(\varphi)$}
\begin{figure*}
  \begin{center}
    \includegraphics[width=15cm]{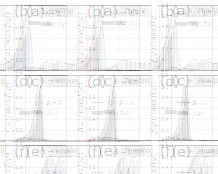}
    \caption{
    Rescaled $f(\varphi)$ and fitting results.
    $(\Phi, N)$ is set as $(\Phi, N)=(0.6, 8192)$ and $Pe$ is varied as
    (a) $Pe=50$, (b) $Pe=100$, (c) $Pe=200$, (d) $Pe=400$, (e) $Pe=800$ and (f) $Pe=1000$, respectively.
    \label{fig: fitting}}
  \end{center}
\end{figure*}

$f(\varphi)$ obtained by actual simulation and the results of fitting $f_g$ and $f_\ell$ to $f(\varphi)$ are shown in Fig. \ref{fig: fitting}.
Figure \ref{fig: fitting} shows that the fitting is generally successful, but there is a problem with the $f(\varphi)$ fitting.
In Figs. \ref{fig: fitting}(b), (c) and (d), the fitting is misaligned with the tail on the left side of the main peak (the peak of high-density side).
Although $f_\ell$ was assumed to be a power-law function that increases with $\varphi$, the distribution may increase more slowly than the power-law one.

\section{A time $\tau$ when the motion switches from glass-like to ballistic}
\begin{figure*}
  \begin{center}
    \includegraphics[width=15cm]{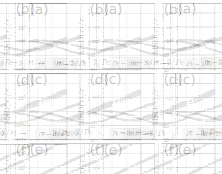}
    \caption{
    The MSD of the particles inside the cluster.
    Each MSD is divided by $\tau$ to make it easier to understand the time when the power-law exponent of $\tau$ switches (from less than $1$ to greater than $1$) in the MSD.
    $(\Phi, N)$ is set as $(\Phi, N)=(0.6, 8192)$ and $Pe$ was changed as
    (a) $Pe=50$, (b) $Pe=100$, (c) $Pe=200$, (d) $Pe=400$, (e) $Pe=800$ and (f) $Pe=1000$, respectively.
    The vertical black dashed lines indicate $\tau=3/Pe$.
    \label{fig: MSDpt}}
  \end{center}
\end{figure*}

The MSD of the particles inside the cluster obtained by the same method as in the main text is shown in Fig. \ref{fig: MSDpt}.
Since each MSD is divided by $\tau$ , the dynamics switches from glass-like to ballistic at the ${\rm MSD(\tau)}/\tau$ minima.
The time when the motion of the particles switches from glass-like to ballistic is about $\tau\sim 3/Pe$ for $Pe\le400$, but much earlier for $Pe\ge800$.
Since the travel distance of particles by the short slip deformation ($\sim1.0$) and the distance traveled by a freely propelled ABP during $\tau=3/Pe$ ($=1.0$) are of similar value, the slip deformation and the time scale $3/Pe$ are supposed to be related, although the details are not clear.

\section{Videos showing particle motion}
Seven videos were created. 
Let us call the seven videos as video 1, video 2,... and video7, respectively\cite{video1, video2, video3, video4, video5, video6, video7}.
The parameters are set to $\Phi=0.6$ and $N=8192$ for all videos.
Videos 1, 2, and 3 show the motion of particles in a cluster.
In one second in the video, the time advances by $5\times(3/Pe)$ in the equation of motion.
The colors of the particles correspond to the indices, white ones for odd numbers and black ones for even numbers.
The $Pe$ values are different for the three videos, 
with $Pe=100$, $Pe=400$, and $Pe=800$ 
set for video1, video2, and video3, respectively.
The video confirms that slip deformations occur within the cluster.

\begin{figure*}
  \begin{center}
    \includegraphics[width=4cm]{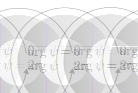}
    \caption{
    The colormap of the argument of the hexatic order parameter.
    \label{fig: hexatic_colormap}}
  \end{center}
\end{figure*}

Videos 4, 5, and 6 show the presence and movement of defects within the cluster.
In one second in the video, the time advances by $5\times(3/Pe)$ in the equation of motion.
The color of the particles corresponds to the argument (of complex number) of the hexatic order parameter described in the following.
Let $\mathcal{N}_i$ be the set of indices of particles near particle $i$, and $\theta_{ij}$ be the angle of the bond vector $\bm{r}_j - \bm{r}_i$ with respect to particle $j\in \mathcal{N}_i$ measured counterclockwise from the positive direction of the $x$-axis.
The hexatic order parameter\cite{PhysRevLett.107.155704, PhysRevLett.114.035702, PhysRevLett.121.098003, D1SM01411K} $\psi_i$ for particle $i$ is defined as 
\begin{align}
  \psi_i = \frac{1}{|\mathcal{N}_i|}\sum_{j\in \mathcal{N}_i}
  e^{i6\theta_{ij}}.
\end{align}
The absolute value of $\psi_i$ indicates the degree of crystallization.
Suppose that the particles form a triangular lattice of crystals.
In this case $|\mathcal{N}_i|=6$ and $\theta_{ij}=\pi n/3,~(n\in \mathbb{N})$ 
so $|\psi_i|$ takes the maximum value 1.
When the crystal is distorted, $|\mathcal{N}_i|\neq 6$ and $\theta_{ij}\neq\pi n/3,~(n\in \mathbb{N})$, $|\psi_i|$ takes a value smaller than $1$.
The argument of $\psi_i$ indicates the inclination of the crystal axis to $x$-axis.
When the entire crystal is gradually rotated by $\pi/3$, the angle $\theta_{ij}$ of each bond vector gradually changes to $\theta_{ij}\to\theta_{ij}+\pi/3$ and $\psi_i$ makes one rotation on the complex number plane.
The colormap of the argument $\arg\psi$ of the hexatic order parameter $\psi$ is shown in Fig. \ref{fig: hexatic_colormap}.
The $Pe$ values are different for the three videos, 
with $Pe=100$, $Pe=400$, and $Pe=800$ 
set for video4, video5, and video6, respectively.
When a defect exists, the coordination number $|\mathcal{N}_i|$ deviates from $6$, and the inclination $\arg\psi$ also deviates from those of the surrounding particles.
Therefore, the color of a particle near the defect deviates from the color of surrounding particles.
The videos show that the dislocations are moving within a cluster.
The moving dislocations are aligned on a curve and form a grain boundary.

\begin{figure*}
  \begin{center}
    \includegraphics[width=7cm]{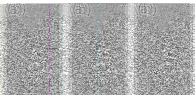}
    \caption{Snapshots of video7.
    (a)0 seconds, (b)30 seconds.
    \label{fig: video_snapshot}}
  \end{center}
\end{figure*}

Video 7 shows the motion of particles in a cluster, similar to videos 1, 2, and 3, but the 
time scale is different, and time advances by $10\times(3/Pe)$ in the equation of motion per second in the video.
The color of the particles corresponds to the indices of particles, white for odd numbers and black for even numbers.
The value of $Pe$ is set to $Pe=400$.
The video confirms that a lot of slip deformation in the cluster result in deformation of the entire cluster.
Snapshots of the video at 0 and 30 seconds are shown in Fig. \ref{fig: video_snapshot}.
The cluster is observed to be circular shape at 0 seconds and deformed into an elongated ellipsoidal shape at 30 seconds.

\section{Number of samples used in the analysis}
The number of samples with the parameter fixed as $\Phi=0.6$ and 
$Pe$ and $N$ are changed is shown in Table \ref{tbl: samples1}.
The number of samples with the parameter fixed as $N=8192$ and 
$Pe$ and $\Phi$ are changed is shown in Table \ref{tbl: samples2}.
\begin{table*}
\small
  \begin{tabular*}{\textwidth}{@{\extracolsep{\fill}}|l||l|l|l|l|l|l|l|l|l|l|l|l|l|l|l|}
  \hline
    ~ & $Pe$= 50 & 60 & 70 & 80 & 90 & 100 & 200 & 300 & 400 & 500 & 600 & 700 & 800 & 900 & 1000 \\ \hline\hline
   $N=1024$ & 15 & 15 & 15 & 15 & 15 &  15 &  10 &  10 &  10 &  10 &  10 &  10 &  10 &  10 &  6\\ \hline
   $2048$   & 15 & 11 & 11 & 10 & 10 &  15 &   8 &   8 &   8 &   8 &   8 &   8 &   8 &   8 &  7\\ \hline
   $4096$   & 15 & 11 & 11 & 10 & 10 &  15 &   5 &   5 &   7 &   7 &   7 &   7 &   7 &   7 & 10\\ \hline
   $8192$   & 15 & 11 & 11 & 10 & 10 &  16 &  10 &  10 &  10 &  10 &  15 &   5 &   5 &  10 & 20\\ 
   \hline
  \end{tabular*}
  \caption{
  The number of samples with the parameter fixed as $\Phi=0.6$ and $Pe$ and $N$ are changed.
  \label{tbl: samples1}}
\end{table*}

\begin{table*}
\small
  \begin{tabular*}{\textwidth}{@{\extracolsep{\fill}}|l||l|l|l|l|l|l|l|l|l|l|l|l|l|l|l|}
  \hline
   ~ & $Pe$=    50 & 60 & 70 & 80 & 90 & 100 & 200 & 300 & 400 & 500 & 600 & 700 & 800 & 900 & 1000 \\ \hline\hline
   $\Phi=0.5$ &  4 &  4 &  4 &  4 &  4 &   4 &   4 &   4 &   4 &   4 &   4 &   4 &   4 &   4 &  4\\ \hline
   $0.6$      & 15 & 11 & 11 & 10 & 10 &  16 &  10 &  10 &  10 &  10 &  15 &   5 &   5 &  10 & 20\\ \hline
   $0.7$      &  4 &  4 &  4 &  4 &  4 &   4 &   4 &   4 &   4 &   4 &   4 &   4 &   8 &   8 &  4\\ \hline
   $0.8$      &  4 &  4 &  4 &  4 &  4 &   4 &   4 &   4 &   4 &   4 &   4 &   4 &   4 &   8 &  8\\ 
   \hline
  \end{tabular*}
  \caption{
  The number of samples with the parameter fixed as $N=8192$ and $Pe$ and $\Phi$ are changed.
  \label{tbl: samples2}}
\end{table*}

\bibliography{references_esi}
\bibliographystyle{rsc}